\documentclass[twocolumn,prb]{revtex4}
\usepackage{amsfonts}
\usepackage[T1]{fontenc}
\usepackage{amsmath,amsbsy,amssymb,graphicx}
\usepackage{times}

\begin{document}

\title{Non-Hermitian non-Abelian topological insulators with $PT$ symmetry}
\author{Motohiko Ezawa}
\affiliation{Department of Applied Physics, University of Tokyo, Hongo 7-3-1, 113-8656,
Japan}

\begin{abstract}
We study a non-Hermitian non-Abelian topological insulator preserving $PT$
symmetry, where the non-Hermitian term represents nonreciprocal hoppings. As
it increases, a spontaneous $PT$ symmetry breaking transition occurs in the
perfect-flat band model from a real-line-gap topological insulator into an
imaginary-line-gap topological insulator. By introducing a band bending
term, we realize two phase transitions, where a metallic phase emerges
between the above two topological insulator phases. We discuss an
electric-circuit realization of non-Hermitian non-Abelian topological
insulators. We find that the spontaneous $PT$ symmetry breaking as well as
the edge states are well observed by the impedance resonance.
\end{abstract}

\maketitle

\section{Introduction}

Topological insulators are one of the most fascinating ideas in contemporary
physics\cite{Hasan,Qi}. They are characterized by topological numbers such
as the winding number, the Chern number and the $\mathbb{Z}_{2}$ index.
However, all of these topological numbers are Abelian.

Non-Abelian topological charges are discussed in three-band models protected
by $PT$ symmetry\cite{WuScience,Tiwari,Guo,YangPRL,Leng} or $C_{2}T$ symmetry\cite{Bouhon,DWang}. They are realized in nodal line semimetals\cite%
{WuScience,Tiwari,YangPRL,Leng,MWang,BJiang,HPark} in three dimensions and
Weyl points\cite{Bouhon} in two dimensions. Non-Abelian topological
insulators in one dimension are studied for three-band models\cite{Guo} and
four-band models\cite{Jiang4}. They are experimentally observed in photonic
systems\cite{YangPRL,DWang}, phononic systems\cite{BJiang,BPeng} and
transmission lines\cite{Guo,Jiang4}. In addition, a generalization to
multi-band theories is proposed in nodal line semimetals\cite{WuScience}. As
far as we aware of, there is no study on non-Hermitian non-Abelian
topological phases so far.

Non-Hermitian topological physics have attracted much attention\cite%
{Bender,Bender2,Malzard,Konotop,Rako,Gana,Zhu,Yao,Jin,Liang,Nori,Lieu,UedaPRX,Coba,Jiang,JPhys,Ashida}. 
In non-Hermitian systems eigenvalues and eigenfunctions are complex in
general. However, they are restricted to be real if $PT$ symmetry is imposed%
\cite{Bender,Mosta,Ruter,Yuce,LFeng,Gana,Weimann}. There is a $PT$ symmetry
breaking transition, where the eigenvalues and eigenfunctions become
complex. Nonreciprocal hopping is such a hopping that the right-going and
left-going hopping amplitude are different\cite{Hatano}. It makes a system
non-Hermitian.

In this paper, we study a non-Hermitian non-Abelian topological insulator in
an $N$ band model with $PT$ symmetry. We show that a spontaneous $PT$
symmetry breaking is induced by increasing the nonreciprocal hoppings from a
phase transition from a real-line-gap topological insulator to an
imaginary-line-gap topological insulator in the case of a perfect-flat band
model. Furthermore, by introducing a band bending term, we may generalize
the model to have a metal with two critical points, where a metallic phase
emerges between the above two topological insulator phases. Finally, we show
how to implement the present model in electric circuits. The edge states and
the spontaneous $PT$ symmetry breaking are well signaled by the impedance
resonance.

\begin{figure}[t]
\centerline{\includegraphics[width=0.49\textwidth]{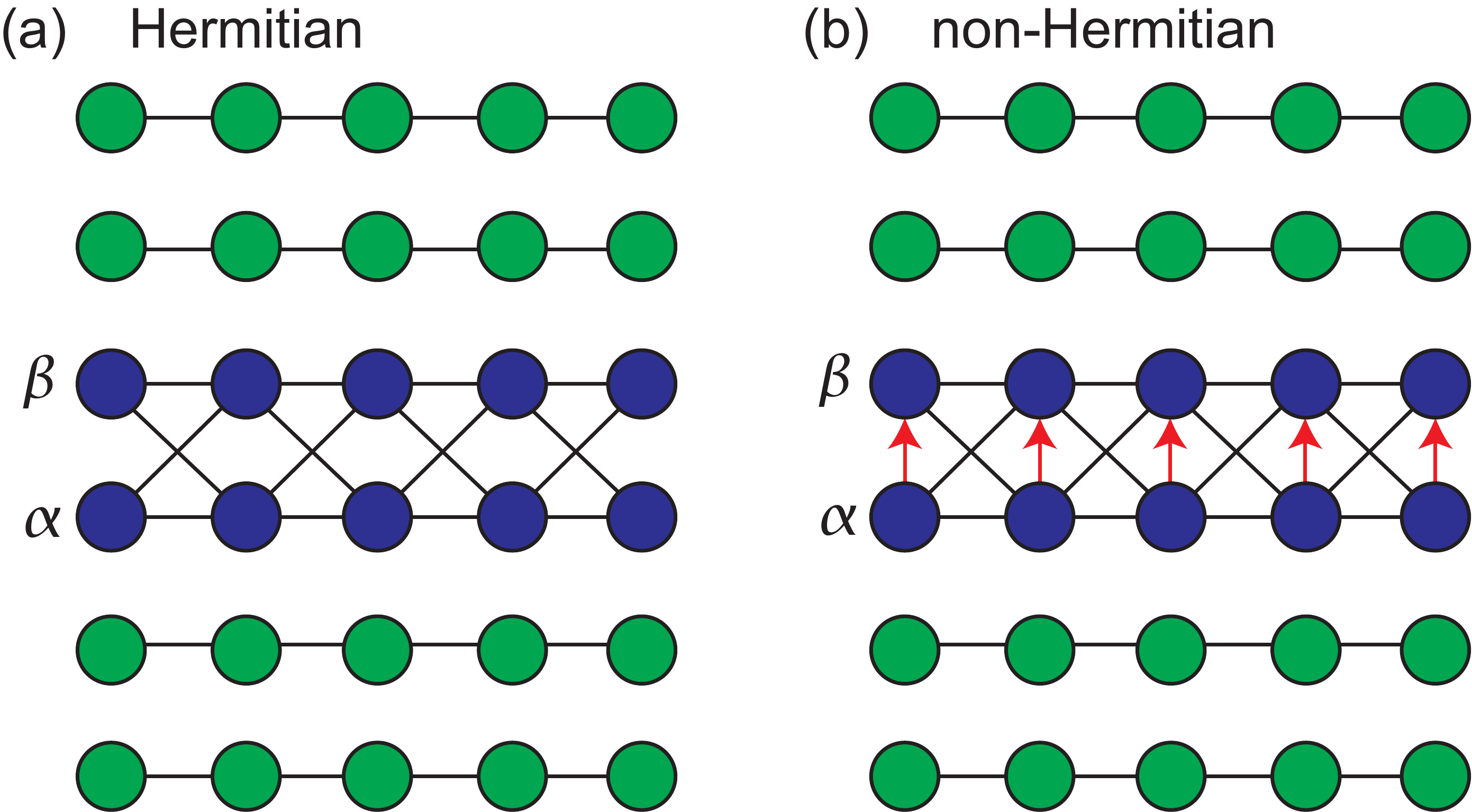}}
\caption{Illustration of the tight-binding Hamiltonian. (a) Hermitian and
(b) non-Hermitian models. Interactions between the $\protect\alpha $ and $\protect\beta $ 
chains yield a non-Abelian topological number. All other
chains shown in green act as spectators. Red arrows represent nonreciprocal
hoppings.}
\label{FigIllust}
\end{figure}

\section{Non-Hermitian Non-Abelian topological insulators}

\subsection{Hermitian Hamiltonian}

\begin{figure*}[t]
\centerline{\includegraphics[width=0.98\textwidth]{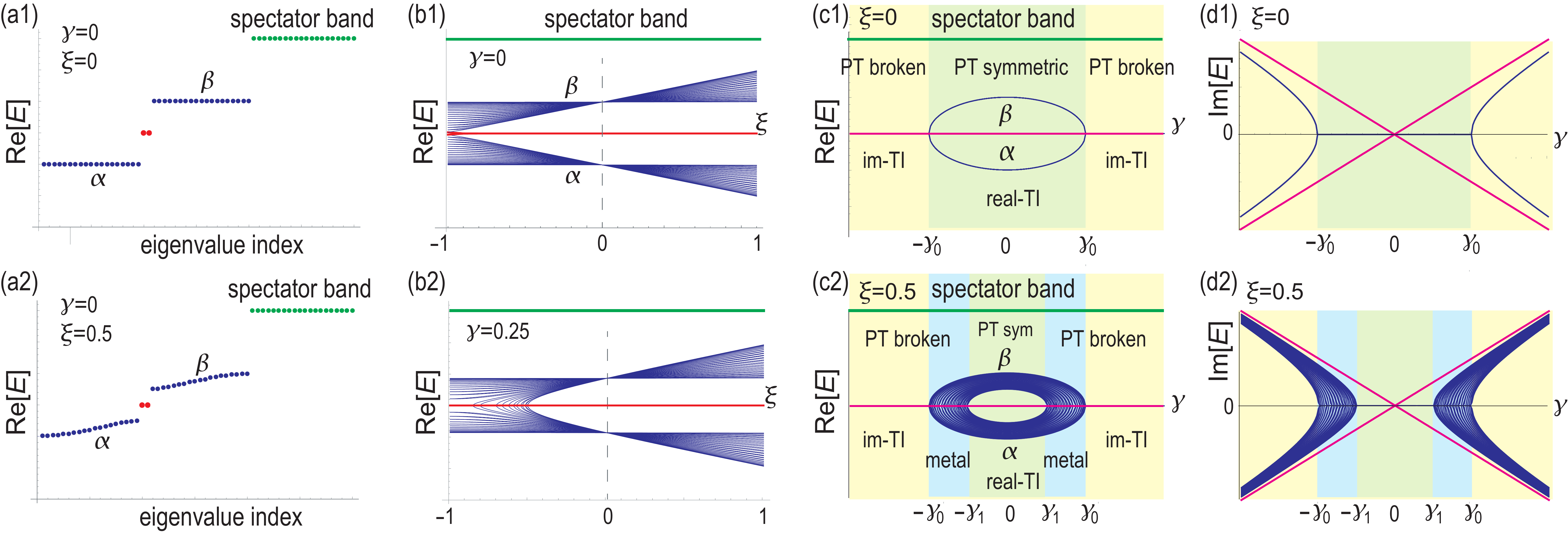}}
\caption{Energy spectrum of the non-Hermitian Hamiltonian in nanoribbon
geometry. Eigenvalues of (a1) the perfectly flat $\protect\alpha $ and 
$\protect\beta $ bands with $\protect\xi =0$ and (a2) the bended bands with 
$\protect\xi =0.5$ shown in blue. The red dots represent the topological edge
states. The band structure as a function of $\protect\xi $ with (b1) 
$\protect\gamma =0$ and (b2) $\protect\gamma =0.25$. The red lines represent
the topological edge states. (c1)
and (c2) Real part of the energy. (d1) and (d2) Imaginary part of the
energy. We have set $\protect\xi =0$ for (c1) and (d1), while we have set 
$\protect\xi =0.5$ for (c2) and (d2). The bulk bands are colored in blue,
while the edge states are colored in red. The spectator band is colored in
green. When $\protect\xi =0$, there are two phases, a real-line-gap
topological insulator (real-TI) phase and an imaginary-line-gap topological
insulator (im-TI) phase. When $\protect\xi \not=0$, a metallic phase emerges
between these two topological insulator phases. We have set $\varepsilon_{\alpha}=1$ and $\varepsilon_{\beta}=2$.}
\label{FigDotBend}
\end{figure*}

We start with a Hermitian system capable to describe a non-Abelian
topological insulator based of the one-dimensional lattice in Fig.\ref{FigIllust}(a). 
We consider generators of $\mathfrak{so}\left( N\right) $
rotation $L_{\alpha \beta }$ indexed by $\alpha $ and $\beta $, whose $ab$
components are defined by 
\begin{equation}
\left( L_{\alpha \beta }\right) _{ab}=\delta _{\alpha b}\delta _{\beta
a}-\delta _{\alpha a}\delta _{\beta b}.  \label{Lab}
\end{equation}%
We consider a $PT$-invariant Hamiltonian in the momentum space given by\cite{WuScience}%
\begin{equation}
H_{\alpha \beta }\left( k\right) =R_{\alpha \beta }\left( \frac{k}{2}\right) 
\text{diag.}\left( \varepsilon _{1},\varepsilon _{2},\cdots ,\varepsilon
_{N}\right) R_{\alpha \beta }\left( \frac{k}{2}\right) ^{\text{t}},
\label{HamilBasic}
\end{equation}%
where $0\leq k<2\pi $, $1\leq \alpha ,\beta \leq N$, $\varepsilon
_{1},\varepsilon _{2},\cdots ,\varepsilon _{N}$ are real, and 
\begin{equation}
R_{\alpha \beta }\left( \frac{k}{2}\right) =e^{\frac{k}{2}L_{\alpha \beta }}
\label{Rot}
\end{equation}%
is a rotation matrix given by%
\begin{align}
\left( R_{\alpha \beta }\left( \frac{k}{2}\right) \right) _{ab}& =\delta
_{ab}+\left( \delta _{a\alpha }\delta _{b\alpha }+\delta _{a\beta }\delta
_{b\beta }\right) \cos \frac{k}{2}  \notag \\
& +\left( \delta _{a\beta }\delta _{b\alpha }-\delta _{a\alpha }\delta
_{b\beta }\right) \sin \frac{k}{2}.
\end{align}%
The Hamiltonian (\ref{HamilBasic}) is explicitly written as%
\begin{align}
H_{\alpha \beta }\left( k\right) & =\frac{\varepsilon _{\alpha }+\varepsilon
_{\beta }}{2}  \notag \\
& +\frac{\varepsilon _{\alpha }-\varepsilon _{\beta }}{2}\left( \delta
_{a\alpha }\delta _{b\alpha }-\delta _{a\beta }\delta _{b\beta }\right) \cos
k  \notag \\
& +\frac{\varepsilon _{\alpha }-\varepsilon _{\beta }}{2}\left( \delta
_{a\beta }\delta _{b\alpha }+\delta _{a\alpha }\delta _{b\beta }\right) \sin
k.
\end{align}%
It is decomposed into two parts,%
\begin{equation}
H_{\alpha \beta }\left( k\right) =\bigoplus_{j\neq \alpha ,\beta
}H_{j}\oplus H_{\alpha \beta }^{\prime }\left( k\right) ,
\end{equation}%
where%
\begin{align}
H_{j}& =\varepsilon _{j}\mathbb{I}_{1}, \\
H_{\alpha \beta }^{\prime }\left( \theta \right) & =\left[ \frac{\varepsilon
_{\alpha }-\varepsilon _{\beta }}{2}\left( 
\begin{array}{cc}
\cos k & \sin k \\ 
\sin k & -\cos k%
\end{array}%
\right) +\frac{\varepsilon _{\alpha }+\varepsilon _{\beta }}{2}\mathbb{I}_{2}\right] .  \label{H'ab}
\end{align}%
The Hamiltonian is nontrivial only for the $\alpha $ and $\beta $ bands,
with eigenvalues $\varepsilon _{\alpha }$ and $\varepsilon _{\beta }$. All
other bands are spectators with respect to the $\alpha $ and $\beta $ bands.
See Fig.\ref{FigIllust}.

The energy spectrum of the bulk Hamiltonian does not change by the rotation (\ref{Rot}) and is given by%
\begin{equation}
E\left( k\right) =\varepsilon _{1},\quad \varepsilon _{2},\quad \cdots
,\quad \varepsilon _{N}.
\end{equation}%
The eigenfunctions for the $2\times 2$ matrix $H_{\alpha \beta }^{\prime
}\left( k\right) $ are%
\begin{align}
\psi _{a}^{+}& =\delta _{a\alpha }\sin \frac{k}{2}+\delta _{a\beta }\cos 
\frac{k}{2}, \\
\psi _{a}^{-}& =-\delta _{a\alpha }\cos \frac{k}{2}+\delta _{a\beta }\sin 
\frac{k}{2},
\end{align}%
while those for $H_{j}$ are $\psi _{a}=\delta _{aj}$.

The $\alpha $ and $\beta $ bands are perfectly flat. They are ($\ell $--$1$)
fold degenerate in a finite chain, where $\ell $ is the number of sites in
the chain. See Fig.\ref{FigDotBend}(a1).

\begin{figure*}[t]
\centerline{\includegraphics[width=0.98\textwidth]{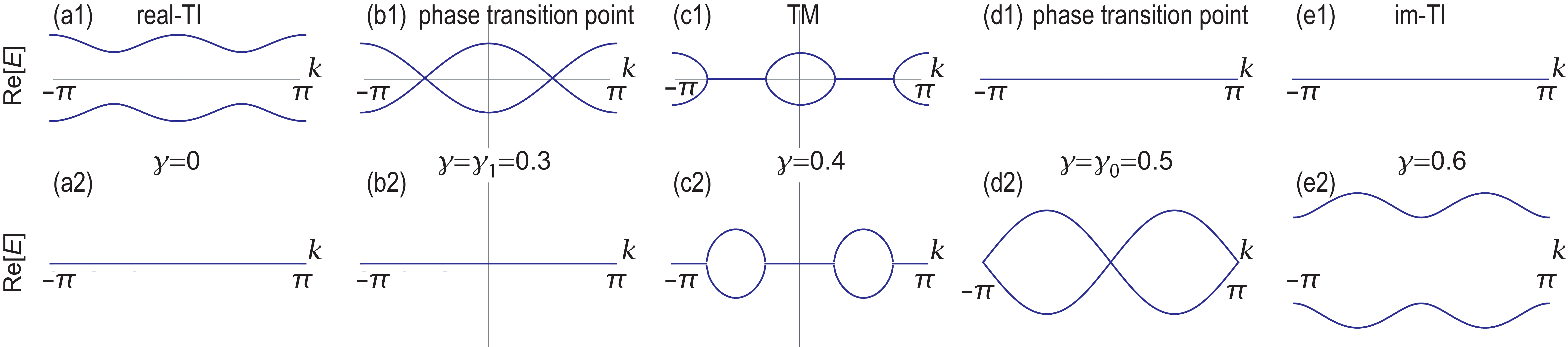}}
\caption{(a1)$\sim $(e1) Real part of the energy and (a2)$\sim $(e2)
imaginary part of the energy. (a1) and (a2) for a real-line-gap topological
insulator (real-TI) with $\protect\gamma =0$; (b1) and (b2) for a phase
transition point with $\protect\gamma =\protect\gamma _{1}=0.3$; (c1) and
(c2) for a metal with $\protect\gamma =0.4$; (d1) and (d2) for a phase
transition point with $\protect\gamma =\protect\gamma _{0}=0.5$; (e1) and
(e2) for an imaginary-line-gap topological insulator (im-TI) with $\protect\gamma =0.6$. 
We have set $\protect\xi =0.2$ for all figures. See also the
caption of Fig.\protect\ref{FigDotBend}. }
\label{FigNHBand}
\end{figure*}

\subsection{Non-Hermitian Hamiltonian}

We generalize the Hermitian non-Abelian system (\ref{HamilBasic}) to a
non-Hermitian non-Abelian system, keeping $PT$ symmetry. We consider the
Hamiltonian%
\begin{equation}
H_{\alpha \beta }^{\prime }\left( k;\gamma ,\xi \right) =H_{\alpha \beta
}^{\prime }\left( k\right) +i\gamma \sigma _{y}+\xi \sigma _{x}\sin k,
\label{NHamil}
\end{equation}%
whose eigenenergies are%
\begin{equation}
E_{\alpha \beta }^{\prime }\left( k;\gamma ,\xi \right) =\frac{\varepsilon
_{\alpha }+\varepsilon _{\beta }\pm \sqrt{g(k;\gamma ,\xi )}}{2},
\end{equation}%
with%
\begin{equation}
g(k;\gamma ,\xi )=\left( \varepsilon _{\alpha }-\varepsilon _{\beta }\right)
^{2}-\gamma ^{2}+4\xi \left( \varepsilon _{\alpha }-\varepsilon _{\beta
}+\xi \right) \sin ^{2}k.
\end{equation}%
We explain the meanings of the $\gamma $ term and the $\xi $ term. The
Hamiltonian (\ref{NHamil}) is Hermitian when $\gamma =0$. When $\xi =0$ in
addition, the band structure is highly degenerate as in Fig.\ref{FigDotBend}(a1). 
This degeneracy is resolved by introducing the $\xi $ term as shown in
Fig.\ref{FigDotBend}(a2). We show the band structure with $\gamma =0$ as a
function of $\xi $ in Fig.\ref{FigDotBend}(b1). The perfect flat bands at 
$\varepsilon _{\alpha }$ and $\varepsilon _{\beta }$ become bended and have
dispersions. We also show the band structure with $\gamma =0.25$ as a
function of $\xi $ in Fig.\ref{FigDotBend}(b2).

We show Re[$E_{\alpha \beta }^{\prime }\left( k;\gamma ,\xi =0\right) $] in
Fig.\ref{FigDotBend}(c1) and Im[$E_{\alpha \beta }^{\prime }\left( k;\gamma
,\xi =0\right) $] in Fig.\ref{FigDotBend}(d1) as a function of $\gamma $.
They are real for $|\gamma |\leq \gamma _{0}$ with%
\begin{equation}
\gamma _{0}=\frac{\varepsilon _{\alpha }-\varepsilon _{\beta }}{2},
\end{equation}%
where $PT$ symmetry is preserved. On the other hand, they are complex for 
$|\gamma |>\gamma _{0}$, and hence $PT$ symmetry is spontaneously broken
there. Namely, although the Hamiltonian is $PT$-symmetric, eigenvalues and
eigenfunctions are no longer real in the spontaneous symmetry broken phase.
We show the real and imaginary parts of the energy as a function of $\gamma $
in Fig.\ref{FigDotBend}(c2) and (c2), where the bulk band has a finite
width. We also show the real and imaginary parts of the energy as a function
of the momentum $k$ in Fig.\ref{FigNHBand}, where the bands have dispersions.

Especially, we have%
\begin{equation}
E_{\alpha \beta }^{\prime }\left( \frac{\pi }{2};\gamma ,\xi \right) =\frac{%
\varepsilon _{\alpha }+\varepsilon _{\beta }\pm \sqrt{h(\gamma ,\xi )}}{2},
\end{equation}%
with%
\begin{equation}
h(\gamma ,\xi )=\left( \varepsilon _{\alpha }-\varepsilon _{\beta }+2\xi
-2\gamma \right) \left( \varepsilon _{\alpha }-\varepsilon _{\beta }+2\xi
+2\gamma \right) .
\end{equation}%
By solving the condition that $E_{\alpha \beta }^{\prime }\left( \frac{\pi }{2};
\gamma ,\xi \right) $ is complex, or $h(\gamma ,\xi )<0$, we find a phase
transition point $\gamma _{1}$ in addition to the phase transition point $\gamma _{0}$ as 
\begin{equation}
\gamma _{1}=\frac{\varepsilon _{\alpha }-\varepsilon _{\beta }}{2}-\xi
=\gamma _{0}-\xi .  \label{EqD}
\end{equation}%
When $\xi >0$, the bulk energy is real for $\left\vert \gamma \right\vert
\leq \gamma _{1}$, complex for $\left\vert \gamma \right\vert >\gamma _{1}$.
On the other hand, when $\xi <0$, the bulk energy is real for $\left\vert
\gamma \right\vert \leq \gamma _{0}$, complex for $\left\vert \gamma
\right\vert >\gamma _{0}$.

\subsection{Tight-binding Hamiltonian}

The tight-binding Hamiltonian (\ref{H'ab}) is written in the coordinate
space as%
\begin{equation}
H_{\alpha \beta }^{\prime }=H_{0}+H_{\gamma }+H_{\xi },  \label{TB}
\end{equation}%
with%
\begin{eqnarray}
H_{0} &=&\frac{\varepsilon _{\alpha }-\varepsilon _{\beta }}{2}%
\sum_{j=1}^{\ell -1}(\left\vert \alpha _{j}\right\rangle \left\langle \alpha
_{j+1}\right\vert +\left\vert \beta _{j}\right\rangle \left\langle \beta
_{j+1}\right\vert ,  \notag \\
&&+i\left\vert \alpha _{j}\right\rangle \left\langle \beta _{j+1}\right\vert
-i\left\vert \beta _{j}\right\rangle \left\langle \alpha _{j+1}\right\vert )+%
\text{h.c.}, \\
H_{\gamma } &=&\gamma \sum_{j=1}^{\ell }(\left\vert \alpha _{j}\right\rangle
\left\langle \beta _{j}\right\vert -\left\vert \beta _{j}\right\rangle
\left\langle \alpha _{j}\right\vert ), \\
H_{\xi } &=&i\xi \sum_{j=1}^{\ell -1}(\left\vert \alpha _{j}\right\rangle
\left\langle \beta _{j+1}\right\vert -\left\vert \beta _{j}\right\rangle
\left\langle \alpha _{j+1}\right\vert )+\text{h.c.},
\end{eqnarray}%
where the first two terms in $H_{0}$ represent normal hoppings, while the
last two terms represent spin-orbit-like imaginary hoppings. The $\xi $
term modifies the spin-orbit-like imaginary hoppings. The $\gamma $ term
represents nonreciprocal hoppings, which make the system non-Hermitian.

The tight-binding Hamiltonians for the spectator bands are simply given by 
\begin{equation}
H_{j\neq \alpha ,\beta }=\sum_{j=1}^{\ell }\varepsilon _{j}\left\vert
j\right\rangle \left\langle j\right\vert +\sum_{j=1}^{\ell
-1}t_{j}\left\vert j\right\rangle \left\langle j+1\right\vert +\text{h.c.},
\end{equation}%
where $\varepsilon _{j}$ is the on-site energy and $t_{j}$ is the hopping
parameter.

In this sense, it is enough to consider only the $\alpha $ and $\beta $
bands for an arbitrary $N$ band system. We illustrate the tight-binding
model in Fig.\ref{FigIllust}.

\begin{figure}[t]
\centerline{\includegraphics[width=0.48\textwidth]{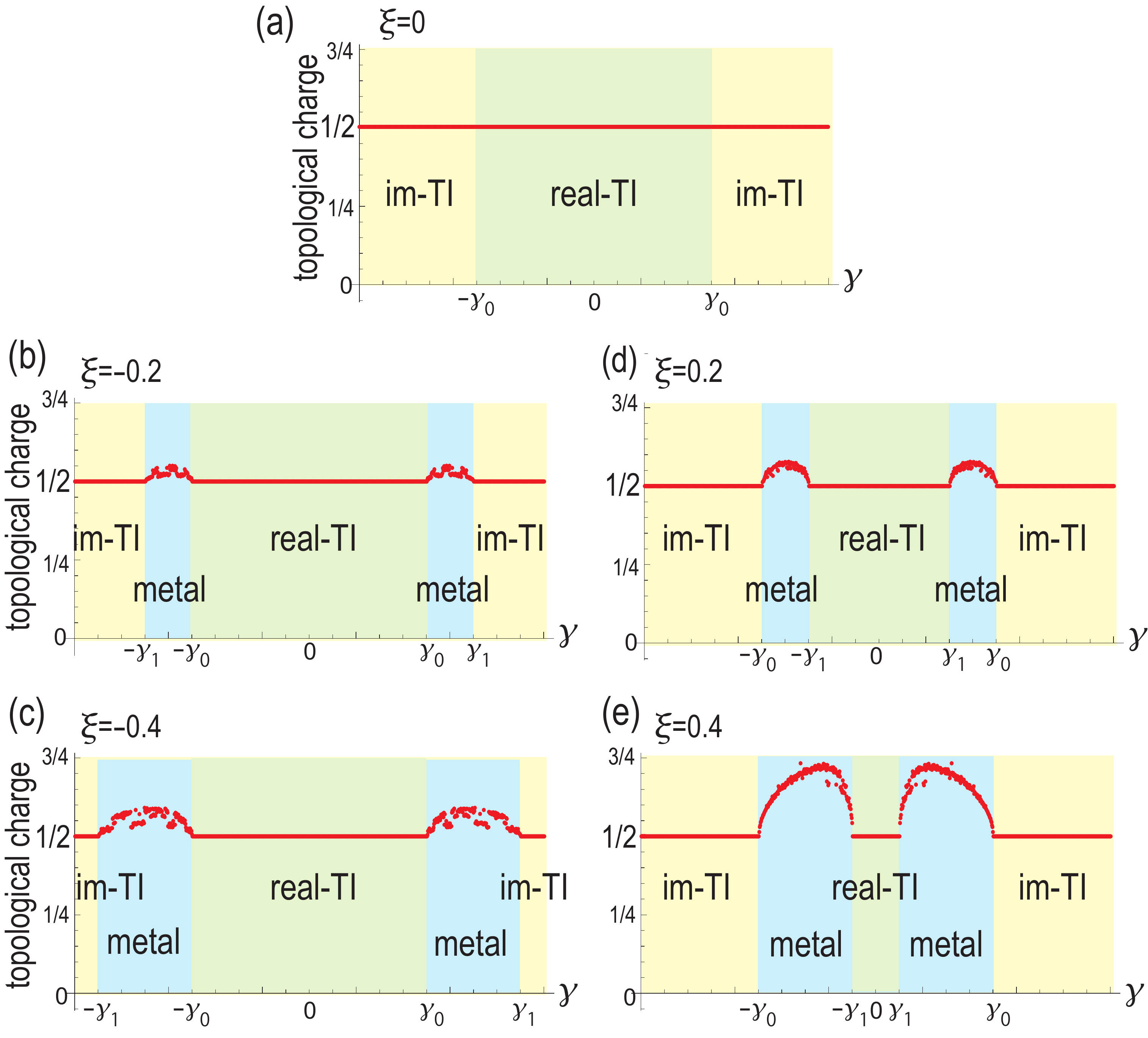}}
\caption{Non-Abelian topological charge marked in red as a function of 
$\protect\gamma $ for various $\protect\xi $. (a) $\protect\xi =0$, 
(b) $\protect\xi =-0.2$, (c) $\protect\xi =-0.4$, (d) $\protect\xi =0.2$ and (e) $\protect\xi =0.4$. 
It is quantized at $1/2$ except for the metallic phase.
In the figures, real-TI (im-TI) stands for real(imaginary)-line-gap
topological insulator phase. }
\label{FigBerry}
\end{figure}

\subsection{Edge states for non-Hermitian model}

We illustrate the tight-binding model (\ref{TB}) in Fig.\ref{FigIllust}(b).
In a finite chain, two localized states emerge at the edges with the energy%
\begin{equation}
E(\xi ,\gamma )=\frac{\varepsilon _{\alpha }+\varepsilon _{\beta }}{2}\pm
i\gamma  \label{EqA}
\end{equation}%
in the presence of the $\gamma $ term and the $\xi $ term. They are
degenerate only in the Hermitian limit ($\gamma =0$). In contrasted to the
bulk band, the eigenenergy (\ref{EqA}) is complex once $\gamma $ is
introduced even for the $PT$ symmetric phase. We show Eq.(\ref{EqA}) as a
function of $\gamma $ in Fig.\ref{FigDotBend}. In contrast to the bulk band,
the eigenenergy (\ref{EqA}) has no $\xi $ dependence: See Figs.\ref{FigDotBend}(d1) and (d2).

When $\xi =0$, the eigenfunctions for the edge states $\psi _{\alpha }\left(
j\right) $ and $\psi _{\beta }\left( j\right) $ at the $j$ site are
perfectly localized at the edges and given by%
\begin{equation}
\psi _{\alpha }\left( j\right) =\frac{1}{\sqrt{2}}\delta _{1,j},\qquad \psi
_{\beta }\left( j\right) =\frac{-i}{\sqrt{2}}\delta _{1,j}  \label{PsiEdgeL}
\end{equation}%
for the left edge, and%
\begin{equation}
\psi _{\alpha }\left( j\right) =\frac{1}{\sqrt{2}}\delta _{\ell ,j},\qquad
\psi _{\beta }\left( j\right) =\frac{i}{\sqrt{2}}\delta _{\ell ,j}
\end{equation}%
for the right edge. Here, $1$ in $\delta _{1,j}$ represent the left edge,
while $\ell $ in $\delta _{\ell ,j}$ represents the right edge. The
perfectly localized edge states for $\xi =0$ are transformed to edge states
with finite penetration depth for $\xi \neq 0$.

\subsection{Non-Hermitian Non-Abelian topological charges}

We define a non-Hermitian non-Abelian Berry connection or a non-Hermitian
Berry-Wilczek-Zee (BWZ) connection by\cite{WZ}%
\begin{equation}
A_{\alpha \beta }^{\text{RL}}\left( \theta \right) =\left\langle \psi
_{\alpha }^{\text{R}}\right\vert \partial _{\theta }\left\vert \psi _{\beta
}^{\text{L}}\right\rangle ,  \label{ABWZ}
\end{equation}%
where%
\begin{equation}
H\left\vert \psi _{\alpha }^{\text{L}}\right\rangle =\varepsilon _{\alpha
}\left\vert \psi _{\alpha }^{\text{L}}\right\rangle
\end{equation}%
is the left eigenfunction, and 
\begin{equation}
H^{\dagger }\left\vert \psi _{\alpha }^{\text{R}}\right\rangle =\varepsilon
_{\alpha }\left\vert \psi _{\alpha }^{\text{R}}\right\rangle
\end{equation}%
is the right eigenfunction.

\begin{figure}[t]
\centerline{\includegraphics[width=0.48\textwidth]{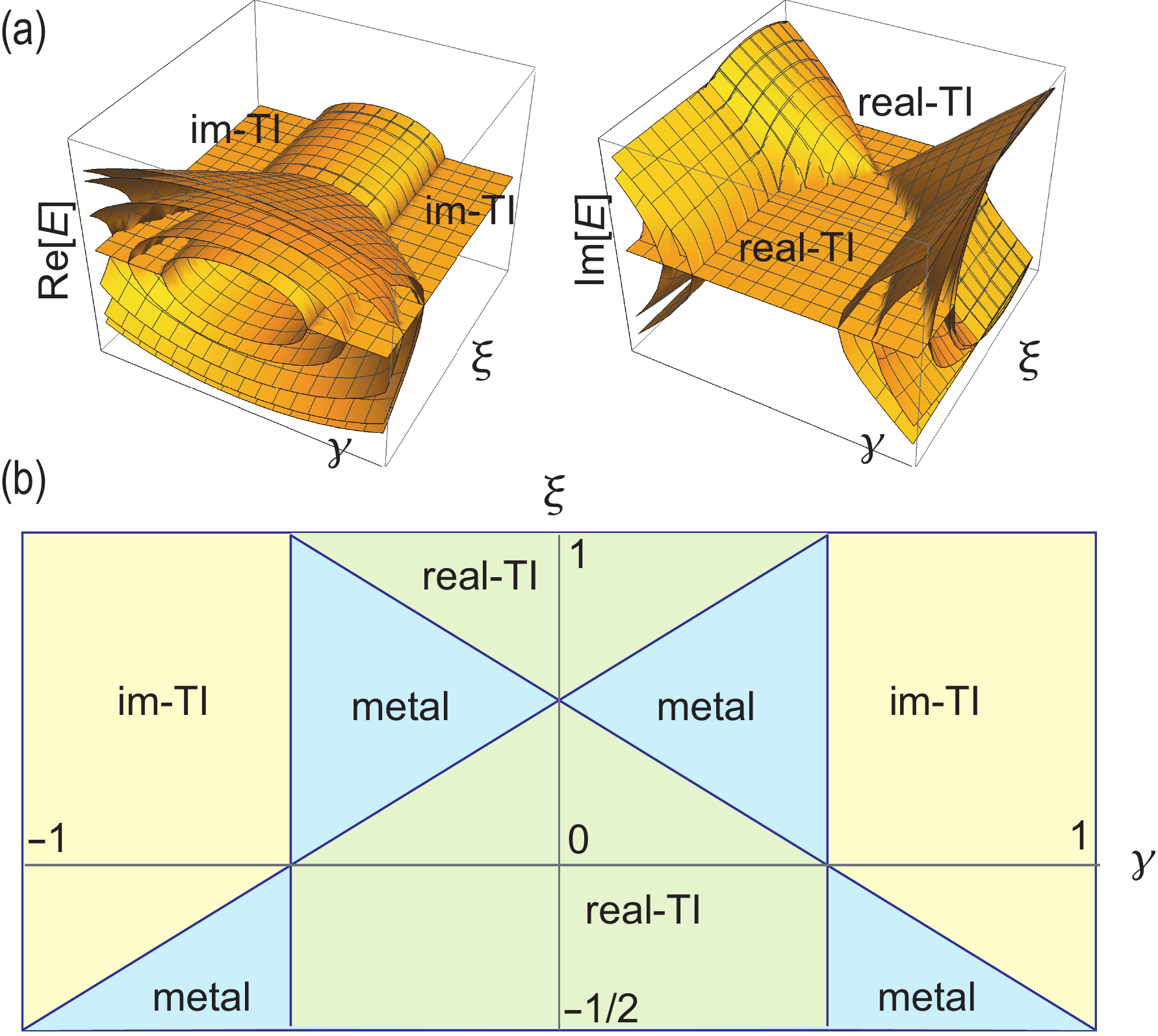}}
\caption{(a) Real and imaginary parts of the bulk-band energy in the 
$(\protect\gamma,\protect\xi)$ plane. (b) Topological phase diagram in the 
$(\protect\gamma,\protect\xi)$ plane. Metallic phase appears except for $\protect\xi =0$. In the figure, real-TI (im-TI) stands for
real(imaginary)-line-gap topological insulator phase. }
\label{FigPhase}
\end{figure}

We define a non-Hermitian BWZ phase by%
\begin{equation}
\Gamma _{\alpha \beta }^{\text{RL}}=\frac{1}{2\pi }\int_{0}^{2\pi }\text{Re}%
\left[ A_{\alpha \beta }^{\text{RL}}\left( \theta \right) \right] d\theta ,
\label{TopCharge}
\end{equation}%
which we use as a non-Hermitian non-Abelian topological charge. The
eigenfunctions are analytically solved as%
\begin{align}
\left\vert \psi _{\pm }^{\text{L}}\right\rangle & =\frac{1}{\sqrt{\left(
\psi _{1}^{\text{L}}\right) ^{2}+\left( \psi _{2}^{\text{L}}\right) ^{2}}}%
\left\{ \psi _{1}^{\text{L}},\psi _{2}^{\text{L}}\right\} ,  \label{PsiL} \\
\psi _{1}^{\text{L}}& =-\gamma _{0}\cos k\pm \sqrt{\gamma _{0}^{2}-\gamma
^{2}+\xi \left( 2\gamma +\xi \right) \sin ^{2}k}, \\
\psi _{2}^{\text{L}}& =\gamma -\left( \gamma _{0}+\xi \right) \sin k,
\end{align}%
and%
\begin{align}
\left\vert \psi _{\pm }^{\text{R}}\right\rangle & =\frac{1}{\sqrt{\left(
\psi _{1}^{\text{R}}\right) ^{2}+\left( \psi _{2}^{\text{R}}\right) ^{2}}}%
\left\{ \psi _{1}^{\text{R}},\psi _{2}^{\text{R}}\right\} ,  \label{PsiR} \\
\psi _{1}^{\text{R}}& =\gamma _{0}\cos k\pm \sqrt{\gamma _{0}^{2}-\gamma
^{2}+\xi \left( 2\gamma +\xi \right) \sin ^{2}k}, \\
\psi _{2}^{\text{R}}& =\gamma +\left( \gamma _{0}+\xi \right) \sin k.
\end{align}

When $\gamma =0$ and $\xi =0$, using (\ref{PsiL}) and (\ref{PsiR}), we
calculate the non-Hermitian BWZ connection (\ref{ABWZ}) numerically and
find that%
\begin{equation}
A_{\alpha \beta }^{\text{RL}}=\frac{1}{2}\left( 
\begin{array}{cc}
0 & -1 \\ 
1 & 0%
\end{array}%
\right) ,  \label{EqE}
\end{equation}%
which leads to the non-Hermitian BWZ phase (\ref{TopCharge}) as 
\begin{equation}
\Gamma _{\alpha \beta }^{\text{RL}}=\frac{1}{2}\left( 
\begin{array}{cc}
0 & -1 \\ 
1 & 0%
\end{array}%
\right) ,  \label{EqF}
\end{equation}%
where we have explicitly written only the nontrivial $2\times 2$ submatrix
within the $N\times N$ matrix. Hence, the topological charges are given by%
\begin{equation}
\Gamma _{\alpha \beta }^{\text{RL}}=-\frac{1}{2}L_{\alpha \beta }
\label{TopoCharge}
\end{equation}%
with Eq.(\ref{Lab}). The topological charges $\Gamma _{\alpha \beta }^{\text{RL}}$ 
obey essentially the same non-Abelian algebra as $L_{\alpha \beta }$.

When $\gamma \neq 0$ and $\xi =0$, we calculate the non-Hermitian BWZ
connection (\ref{ABWZ}) to find that it is no longer a constant. However, the
non-Hermitian BWZ phase (\ref{TopCharge}) is calculated as in Eq.(\ref{EqF}), 
and hence the topological charge is quantized as in Eq.(\ref{TopoCharge})
for any $\gamma $. Nevertheless, the eigenfunctions as well as the
eigenvalues are real (i.e., real-line-gap topological insulator phase) only
for $\gamma ^{2}\leq \gamma _{0}^{2}$, while the eigenvalues and the
eigenfunctions are complex (i.e. imaginary-line-gap topological insulator
phase) for $\gamma ^{2}>\gamma _{0}^{2}$. Hence, PT symmetry is preserved
only for $\gamma ^{2}\leq \gamma _{0}^{2}$, and it is spontaneously broken
for $\gamma ^{2}>\gamma _{0}^{2}$. The system undergoes a phase transition
at $\gamma =\pm \gamma _{0}$.

When $\gamma \neq 0$ and $\xi \neq 0$, we have numerically calculated the
topological charge (\ref{TopCharge}) with the use of (\ref{PsiL}) and (\ref{PsiR}). 
We have shown the $\left( 2,1\right) $ component of the $2\times 2$
matrix $\Gamma _{\alpha \beta }^{\text{RL}}$ for various values of $\xi $ in
Fig.\ref{FigBerry}. It is quantized to be 1/2 for $\gamma ^{2}\leq \gamma
_{1}^{2}$ and $\gamma ^{2}>\gamma _{0}^{2}$ when $\xi >0$, while $\gamma
^{2}\leq \gamma _{0}^{2}$ and $\gamma ^{2}>\gamma _{1}^{2}$ when $\xi <0$,
where $\gamma _{1}=\gamma _{0}-\xi $ as in Eq.(\ref{EqD}). On the other
hand, it is not quantized for the metallic phase that emerges between 
$\gamma _{0}$ and $\gamma _{1}$, as in Fig.\ref{FigBerry}. It is concluded
that the topological charges are quantized and given by Eq.(\ref{TopoCharge}) in the insulator phases.

\subsection{Topological phase diagram}

In non-Hermitian systems, there are line-gap insulators and point-gap
insulators\cite{UedaPRX,Kawabata} in general. In the point-gap insulator,
there is a gap in $|E|$. On the other hand, there are two types of line-gap
insulators. A real-line gap topological insulator has a gap in Re$\left[E\right] $, 
while an imaginary-line-gap topological insulator has a gap in Im$\left[ E\right] $. 
We first consider the case $\xi >0$. For $\left\vert
\gamma \right\vert <\gamma _{1}$, the system is a non-Hermitian line-gap
topological insulator along the Re$\left[ E\right] $. The systems is
metallic for $\gamma _{1}\leq \left\vert \gamma \right\vert \leq \gamma _{0}$. 
For $\left\vert \gamma \right\vert >\gamma _{0}$, the system is a
non-Hermitian line-gap topological insulator along the Im$\left[ E\right] $.
If $\xi <0$, the system is a real-line-gap topological insulator for 
$\left\vert \gamma \right\vert <\gamma _{0}$, it is a metal for $\gamma
_{0}\leq \left\vert \gamma \right\vert \leq \gamma _{1}$ and it is an
imaginary-line-gap topological insulator for $\left\vert \gamma \right\vert
>\gamma _{1}$. We show the topological phase diagram in Fig.\ref{FigPhase}(b). 
It is consistent with the real and imaginary parts of the energy in the 
$\gamma $-$\xi $ plane as shown in Fig.\ref{FigPhase}(a).

\begin{figure}[t]
\centerline{\includegraphics[width=0.49\textwidth]{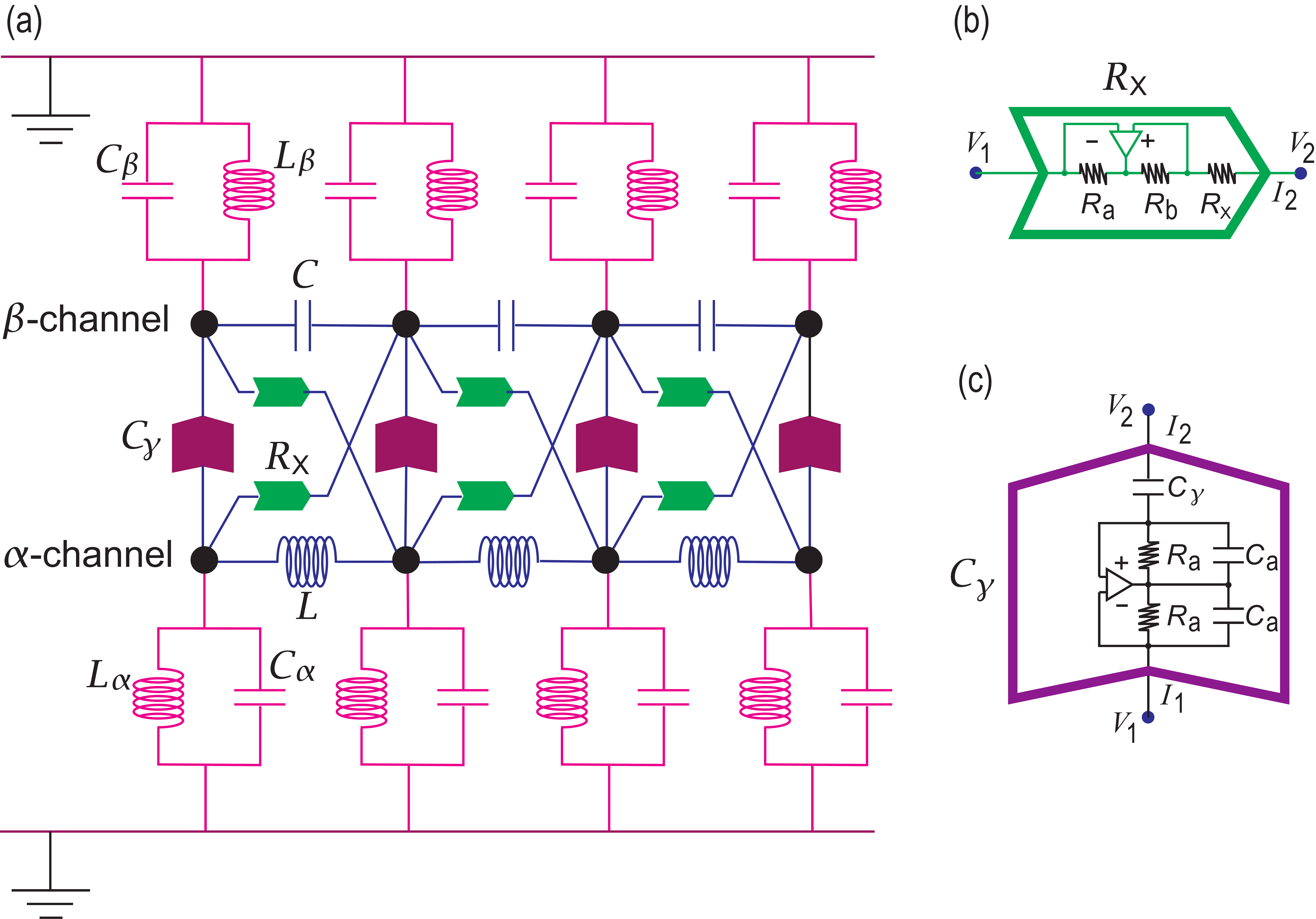}}
\caption{(a) Illustration of the electric circuit corresponding to the
lattice in Fig.\protect\ref{FigIllust}(b). The hopping along the $\protect%
\alpha $-chain ($\protect\beta $-chain) is represented by the inductance $L$ (the capacitance $C$). (b) Negative
impedance converter $R_{X}$ represents an imaginary hopping\protect\cite{Hofmann}. 
(c) Operational amplifier circuit $C_{\protect\gamma }$
represents a nonreciprocal hopping\protect\cite{NonR}.}
\label{FigOpeCircuit}
\end{figure}

\begin{figure*}[t]
\centerline{\includegraphics[width=0.99\textwidth]{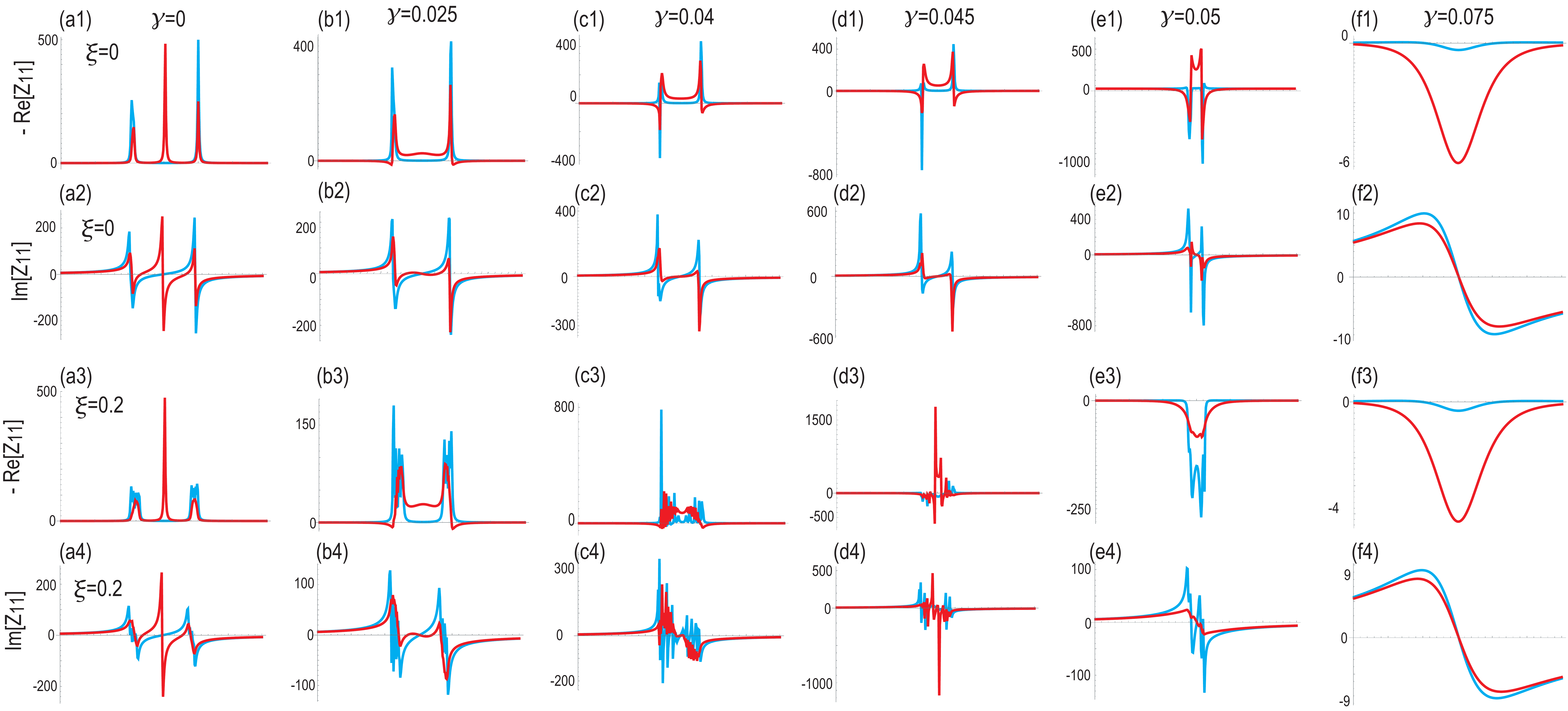}}
\caption{Real and imaginary parts of impedance $Z_{aa}$ at the edge as a
function of frequency $\protect\omega $. (a1) $\sim $(a4) Hermitian model with $\protect\gamma =0$. 
(b1)$\sim $(b4) Non-Hermitian model with $\protect\gamma =0.025$, 
(c1)$\sim $(c4) with $\gamma =0.04$ (phase transition point  $\gamma_1$), (d1)$\sim $(d5) with $\protect\gamma =0.045$, (e1)$\sim $(e4) with $\protect\gamma =0.05$ (phase
transition point $\gamma_0$), and (f1)$\sim $(f4) 
with $\protect\gamma =0.075$. We have used a finite chain with open boundary
condition (red) and periodic boundary condition (cyan). (a1)$\sim $(f2) $\protect\xi =0$. 
(a3)$\sim $(f4) $\protect\xi =0.2$. We have set $\varepsilon_{\alpha}=1$ and $\varepsilon_{\beta}=1.1$. The length of the chain
is 20.}
\label{FigImpe}
\end{figure*}

\section{Electric circuit simulation}

An electric circuit is described by the Kirchhoff current law. By making the
Fourier transformation with respect to time, the Kirchhoff current law is
expressed as 
\begin{equation}
I_{a}\left( \omega \right) =\sum_{b}J_{ab}\left( \omega \right) V_{b}\left(
\omega \right) ,
\end{equation}%
where $I_{a}$ is the current between node $a$ and the ground, while $V_{b}$
is the voltage at node $b$. The matrix $J_{ab}\left( \omega \right) $ is
called the circuit Laplacian. Once the circuit Laplacian is given, we can
uniquely setup the corresponding electric circuit. By equating it with the
Hamiltonian $H$ as\cite{TECNature,ComPhys} 
\begin{equation}
J_{ab}\left( \omega \right) =i\omega H_{ab}\left( \omega \right) ,
\label{CircuitLap}
\end{equation}%
it is possible to simulate various topological phases of the Hamiltonian by
electric circuits\cite%
{TECNature,ComPhys,Hel,Lu,EzawaTEC,EzawaLCR,Research,Zhao,EzawaSkin,Garcia,Hofmann,EzawaMajo,Tjunc,NonR}. 
The relations between the parameters in the Hamiltonian and in the
electric circuit are determined by this formula.

In the present problem, only the $\alpha $-chain and the $\beta $-chain are
active in the tight-binding Hamiltonian as in Fig.\ref{FigIllust}. Thus, we
need only a $2\times 2$ matrix. The circuit Laplacian follows from the
Hamiltonian (\ref{NHamil}) as%
\begin{equation}
J_{\alpha \beta }^{\prime }\left( k\right) =i\omega \left[ \left( 
\begin{array}{cc}
-\frac{L}{\omega ^{2}}\cos k & f_{+} \\ 
f_{-} & C\cos k%
\end{array}%
\right) +\frac{\varepsilon _{\alpha }+\varepsilon _{\beta }}{2}\mathbb{I}_{2}\right] ,  \label{EqB}
\end{equation}%
with%
\begin{equation}
f_{\pm }=\frac{1}{\omega R_{X}}\left( 1+\xi \right) \sin k\pm \gamma .
\label{EqC}
\end{equation}%
We may design the electric circuit to realize this circuit Laplacian as in
Fig.\ref{FigOpeCircuit}. The main part consists of the $\alpha $-channel and
the $\beta $-channel corresponding to the $\alpha $-chain and the $\beta $-chain 
in the lattice in Fig.\ref{FigIllust}. Additionally, each node in the 
$i$-channel is connected to the ground via a set of inductor $L_{i}$ and
capacitor $C_{i}$, where $i=\alpha $ or $\beta $, in order to realize the
diagonal term $\varpropto (\varepsilon _{\alpha }+\varepsilon _{\beta })$ in
Eq.(\ref{EqB}).

Hopping terms along the $\alpha $-chain and the $\beta $-chain are\
described by the diagonal terms in Eq.(\ref{EqB}), where $\pm \cos k=\pm
(e^{ik}+e^{-ik})/2$ represents the plus (minus) hopping in the tight-bind
model. To simulate the positive and negative hoppings in the Hamiltonian, we
replace them with the capacitance $i\omega C$ and the inductance $1/i\omega
L $, respectively.

Hopping terms across the $\alpha $-chain and the $\beta $-chain are\
described by the off-diagonal terms $f_{\pm }$ in Eq.(\ref{EqB}), which
consist of two terms proportional to $\sin k$ and $\gamma $.

(i) The term proportional to $\sin k$ produces the cross hopping, where 
$\sin k=(e^{ik}-e^{-ik})/2i$ represents an imaginary hopping in the
tight-bind model. The imaginary hopping is implemented by a negative
impedance converter $R_{X}$ with current inversion\cite{Hofmann}, as is
constructed based on an operational amplifier with resistors: See Fig.\ref{FigOpeCircuit}(b). 
The voltage-current relation is given by 
\begin{equation}
\left( 
\begin{array}{c}
I_{1} \\ 
I_{2}%
\end{array}%
\right) =\frac{1}{R_{X}}\left( 
\begin{array}{cc}
-\nu & \nu \\ 
-1 & 1%
\end{array}%
\right) \left( 
\begin{array}{c}
V_{1} \\ 
V_{2}%
\end{array}%
\right) ,
\end{equation}%
with $\nu =R_{b}/R_{a}$, where $R_{X}$, $R_{a}$ and $R_{b}$ are the
resistances in an operational amplifier. We note that the resistors in the
operational amplifier circuit are tuned to be $\nu =1$ in the literature\cite{Hofmann} 
so that the system becomes Hermitian, where the corresponding
Hamiltonian represents a spin-orbit interaction. It produces the Hamiltonian%
\begin{equation}
H=\frac{1}{\omega R_{X}}\left( 
\begin{array}{cc}
i & -i \\ 
i & -i%
\end{array}%
\right)
\end{equation}%
for the Hermitian limit.

(ii) The term $\varpropto \gamma $ produces the nonreciprocal hopping terms,
which are vertical hoppings represented by red arrows in Fig.\ref{FigIllust}(b). 
The nonreciprocal hopping is constructed by a combination of an
operational amplifier and capacitors\cite{NonR},%
\begin{equation}
\left( 
\begin{array}{c}
I_{ij} \\ 
I_{ji}%
\end{array}%
\right) =i\omega C_{\gamma }\left( 
\begin{array}{cc}
-1 & 1 \\ 
-1 & 1%
\end{array}%
\right) \left( 
\begin{array}{c}
V_{i} \\ 
V_{j}%
\end{array}%
\right) ,
\end{equation}%
as in Fig.\ref{FigOpeCircuit}(c). It corresponds to the Hamiltonian%
\begin{equation}
H=C_{\gamma }\left( 
\begin{array}{cc}
-1 & 1 \\ 
-1 & 1%
\end{array}%
\right) .
\end{equation}

In this way, the tight-binding Hamiltonian for the present non-Hermitian
non-Abelian topological system is implemented in the electric circuit given
in Fig.\ref{FigOpeCircuit}.

\subsection{Impedance resonance}

The band structure as well as edge states are well observed by impedance
resonance, which is defined\cite{TECNature,ComPhys,Hel} by 
\begin{equation}
Z_{ab}=V_{a}/I_{b}=G_{ab},
\end{equation}%
where $G=J^{-1}$ is the Green function. Taking the nodes $a=b$ at an edge,
we show the real and imaginary parts of the impedance for a finite chain as
a function of $\omega $ in Fig.\ref{FigImpe}, which are marked in red. For
comparison, we also show the impedance for a periodic boundary condition in
cyan, where the edge states are absent.

We first study the Hermitian case ($\gamma =0$) with $\xi =0$, where the
impedance is shown in Fig.\ref{FigImpe}(a1) and (a2). The edge impedance
resonance is clear by comparing the periodic boundary condition and the open
boundary condition. There are only two bulk peaks in cyan at Re[$E_{\alpha
\beta }^{\prime }\left( k;\gamma ,\xi \right) $]. On the other hand, there
is an additional peak in red due to the edge states between two bulk peaks,
as corresponds to Fig.\ref{FigDotBend}(a1).

Next, we show the impedance for various nonreciprocity $\gamma $ with $\xi =0
$ in Fig.\ref{FigImpe}(a1)$\sim $(f1) and (a2)$\sim $(f2).
The edge impedance resonance rapidly decreases as
the increase of $\gamma $, as shown in Fig.\ref{FigImpe}(b1). This is due to
the imaginary contribution in Eq.(\ref{EqA}). Then, the distance between two
bulk peaks becomes narrower, which is consistent with Re[$E_{\alpha \beta
}^{\prime }\left( k;\gamma ,\xi =0\right) $] as shown in Fig.\ref{FigDotBend}(c1). 
The two bulk peaks merge into one peak at the spontaneous $PT$
symmetry breaking point $\gamma _{0}$, as shown in Fig.\ref{FigImpe}(e1).
The bulk impedance resonance is very strong due to the gap closing of the
bulk band. We also observe the edge impedance resonance in the
imaginary-line gap topological insulating phase, where the impedance
resonance is weak comparing to Fig.\ref{FigImpe}(a1) as shown in Fig.\ref{FigImpe}(f1). 
This is also the imaginary contribution in Eq.(\ref{EqA}).

We also show the impedance for finite $\xi $ in Fig.\ref{FigImpe}(a3)$\sim $(f3) and (a4)$\sim $(f4),  
as corresponds to Fig.\ref{FigDotBend}(c2). 
The bulk impedance peaks become broad, which
reflects the broadening of the bulk bands. As a result, the edge impedance
peak becomes clearer as in Fig.\ref{FigImpe}(a3) in comparison to Fig.\ref{FigImpe}(a1). 
There are strong cyan resonances at the phase transition
point $\gamma _{1}$ point as shown in Fig.\ref{FigImpe}(c3) and (c4). It is
due to the gap closing of the bulk band. In Fig.\ref{FigImpe}(d3) and (d4),
the impedance structure is complicated, which reflects the metallic band
structure. The effect of the $\xi $ term is negligible for the
imaginary-line-gap topological phase as shown in Fig.\ref{FigImpe}(f3) and
(f4) since the peak of the impedance is broad even for $\xi =0$ in Fig.\ref{FigImpe}(f1) and (f2). 
Here, note that $\xi $ appears only in the form of $(1+\xi )$ in Eq.(\ref{EqC}).

\section{Conclusion}

We have proposed a non-Hermitian non-Abelian topological insulator model by
imposing $PT$ symmetry in one dimension. It describes a real-line-gap
topological insulator with real eigenvalues in the Hermitian limit. The
system undergoes a spontaneous breakdown of $PT$ symmetry as the
non-Hermitian term increases, and turns out to describe an
imaginary-line-gap topological insulator, when the bulk bands are perfectly
flat. When we introduce a bulk bending term, there are two phase transitions
with the emergence of a metal with complex eigenvalues between the above two
topological insulators. 
Finally, we have presented how to construct
these models in electric circuits. We have shown that 
the spontaneous $PT$ symmetry breaking as well as topological edge states are well signaled by
measuring the frequency dependence of the impedance.

\section*{Acknowledgement}

The author is very much grateful to N. Nagaosa for helpful discussions on
the subject. This work is supported by the Grants-in-Aid for Scientific
Research from MEXT KAKENHI (Grants No. JP17K05490 and No. JP18H03676). This
work is also supported by CREST, JST (JPMJCR16F1 and JPMJCR20T2).


\begin{thebibliography}{99}
\bibitem{Hasan} M. Z. Hasan and C. L. Kane, Rev. Mod. Phys. \textbf{82},
3045 (2010).

\bibitem{Qi} X.-L. Qi and S.-C. Zhang, Rev. Mod. Phys. \textbf{83}, 1057
(2011).

\bibitem{WuScience} Q. Wu, A. A. Soluyanov, T. Bzdusek, Science \textbf{365}, 1273 (2019)

\bibitem{Tiwari} A. Tiwari and T. Bzdu\v{s}ek, Phys. Rev. B \textbf{101},
195130 (2020)

\bibitem{YangPRL} E. Yang, B. Yang, O. You, H.-c. Chan, P. Mao, Q. Guo, S.
Ma, L. Xia, D. Fan, Y. Xiang, S. Zhang, Phys. Rev. Lett. \textbf{125},
033901 (2020)

\bibitem{Leng} P. M. Lenggenhager, X. Liu, S. S. Tsirkin, T. Neupert and T.
Bzdusek, arXiv:2008.02807

\bibitem{Guo} Q. Guo, T. Jiang, R.-Y. Zhang, L. Zhang, Z.-Q. Zhang, B. Yang,
S. Zhang, C. T. Chan, Nature 594, 195 (2021)

\bibitem{Bouhon} A. Bouhon, Q. Wu, R.-J. Slager, H. Weng, O. V. Yazyev and
T. Bzdusek, Nature Physics \textbf{16}, 1137 (2020)

\bibitem{DWang} D. Wang, B. Yang, Q. Guo, R.-Y. Zhang, L. Xia, X. Su, W.-J.
Chen, J. Han, S. Zhang, C. T. Chan, Light: Science \& Applications \textbf{10}, 83 (2021)

\bibitem{HPark} H. Park, S. Wong, X. Zhang, S. S. Oh, arXiv:2102.12546

\bibitem{BJiang} Bin Jiang, Adrien Bouhon, Zhi-Kang Lin, Xiaoxi Zhou, Bo
Hou, Feng Li, Robert-Jan Slager, Jian-Hua Jiang, arXiv:2104.13397

\bibitem{MWang} M. Wang, S. Liu, Q. Ma, R.-Y. Zhang, D. Wang, Q. Guo, B.
Yang, M. Ke, Z. Liu, and C. T. Chan, arXiv:2106.06711

\bibitem{Jiang4} T. Jiang, Q. Guo, R.-Y. Zhang, Z.-Q. Zhang, B. Yang, C. T.
Chan, arXiv:2106.16080

\bibitem{BPeng} B. Peng, A. Bouhon, B. Monserrat, R.-J. Slager,
arXiv:2105.08733

\bibitem{Bender} C. M. Bender and S. Boettcher, Phys. Rev. Lett. \textbf{80}, 5243 (1998).

\bibitem{Gana} R. El-Ganainy, K. G. Makris, M. Khajavikhan, Z. H.
Musslimani, S. Rotter and D. N. Christodoulides, Nat. Physics \textbf{14},
11 (2018).

\bibitem{Bender2} C. M. Bender, D. C. Brody, and H. F. Jones, Phys. Rev.
Lett. \textbf{89}, 270401 (2002).

\bibitem{Malzard} S. Malzard, C. Poli, H. Schomerus, Phys. Rev. Lett. 
\textbf{115}, 200402 (2015).

\bibitem{Konotop} V. V. Konotop, J. Yang, and D. A. Zezyulin, Rev. Mod.
Phys. \textbf{88}, 035002 (2016).

\bibitem{Rako} T. Rakovszky, J. K. Asboth, and A. Alberti, Phys. Rev. B 
\textbf{95}, 201407(R) (2017).

\bibitem{Zhu} B. Zhu, R. Lu and S. Chen, Phys. Rev. A \textbf{89}, 062102
(2014).

\bibitem{Yao} S. Yao and Z. Wang, Phys. Rev. Lett. \textbf{12}1, 086803
(2018).

\bibitem{Jin} L. Jin and Z. Song, Phys. Rev. B \textbf{99}, 081103 (2019).

\bibitem{Liang} S.-D. Liang and G.-Y. Huang, Phys. Rev. A \textbf{87},
012118 (2013).

\bibitem{Nori} D. Leykam, K. Y. Bliokh, Chunli Huang, Y. D. Chong, and
Franco Nori, Phys. Rev. Lett. \textbf{118}, 040401 (2017).

\bibitem{Lieu} S. Lieu, Phys. Rev. B \textbf{97}, 045106 (2018).

\bibitem{UedaPRX} Z. Gong, Y. Ashida, K. Kawabata, K. Takasan, S.
Higashikawa and M. Ueda, Phys. Rev. X \textbf{8}, 031079 (2018).

\bibitem{Coba} E. Cobanera, A. Alase, G. Ortiz, L. Viola, Phys. Rev. B 
\textbf{98}, 245423 (2018).

\bibitem{Jiang} H. Jiang, C. Yang and S. Chen, Phys. Rev. A \textbf{98},
052116 (2018)

\bibitem{JPhys} A. Ghatak, T. Das, J. Phys.: Condens. Matter \textbf{31},
263001 (2019).

\bibitem{Ashida} Y. Ashida, Z. Gong, M. Ueda, Advances in Physics \textbf{69}, 3 (2020)

\bibitem{Mosta} A. Mostafazadeh, J. Math. Phys \textbf{43}, 205 (2002).

\bibitem{Ruter} C. E. Ruter, K. G. Makris, R. El-Ganainy, D. N.
Christodoulides, M. Segev and D.Kip, Nat. Phys. \textbf{6}, 192 (2010)

\bibitem{Yuce} C. Yuce, Physics Letters A \textbf{379}, 12213 (2015).

\bibitem{LFeng} L. Feng, R. El-Ganainy and L. Ge, Nature Photonics \textbf{11}, 752 (2017).

\bibitem{Weimann} S. Weimann, M. Kremer, Y. Plotnik, Y. Lumer, S. Nolte, K.
G. Makris, M. Segev, M. C. Rechtsman and A. Szameit, Nature Materials 
\textbf{16}, 433 (2017)

\bibitem{Hatano} N. Hatano and D. R. Nelson, Phys. Rev. Lett. \textbf{77},
570 (1996): Phys. Rev. B \textbf{56}, 8651 (1997): Phys. Rev. B \textbf{58},
8384 (1998).

\bibitem{WZ} F. Wilczek and A. Zee Phys. Rev. Lett. \textbf{52}, 2111 (1984)

\bibitem{Kawabata} K. Kawabata, K. Shiozaki, M. Ueda, and M. Sato, Phys.
Rev. X \textbf{9}, 041015 (2019).

\bibitem{TECNature} S. Imhof, C. Berger, F. Bayer, J. Brehm, L. Molenkamp,
T. Kiessling, F. Schindler, C. H. Lee, M. Greiter, T. Neupert, R. Thomale,
Nat. Phys. \textbf{14}, 925 (2018).

\bibitem{ComPhys} C. H. Lee , S. Imhof, C. Berger, F. Bayer, J. Brehm, L. W.
Molenkamp, T. Kiessling and R. Thomale, Communications Physics, \textbf{1},
39 (2018).

\bibitem{Hel} T. Helbig, T. Hofmann, C. H. Lee, R. Thomale, S. Imhof, L. W.
Molenkamp and T. Kiessling, Phys. Rev. B 99, 161114 (2019).

\bibitem{Lu} Y. Lu, N. Jia, L. Su, C. Owens, G. Juzeliunas, D. I. Schuster
and J. Simon, Phys. Rev. B \textbf{99}, 020302 (2019).

\bibitem{Research} K. Luo, R. Yu and H. Weng, Research (2018), ID 6793752.

\bibitem{Zhao} E. Zhao, Ann. Phys. \textbf{399}, 289 (2018).

\bibitem{EzawaTEC} M. Ezawa, Phys. Rev. B \textbf{98}, 201402(R) (2018).

\bibitem{Garcia} M. Serra-Garcia, R. Susstrunk and S. D. Huber, Phys. Rev. B 
\textbf{99}, 020304 (2019).

\bibitem{Hofmann} T. Hofmann, T. Helbig, C. H. Lee, M. Greiter, R. Thomale,
Phys. Rev. Lett. \textbf{122}, 247702 (2019).

\bibitem{EzawaMajo} M. Ezawa, Phys. Rev. B \textbf{100}, 045407 (2019)

\bibitem{Tjunc} M. Ezawa, Phys. Rev. B \textbf{102}, 075424 (2020)

\bibitem{EzawaLCR} M. Ezawa, Phys. Rev. B \textbf{99}, 201411(R) (2019).

\bibitem{EzawaSkin} M. Ezawa, Phys. Rev. B \textbf{99}, 121411(R) (2019).

\bibitem{NonR} T. Helbig, T. Hofmann, S. Imhof, M. Abdelghany, T. Kiessling,
L. W. Molenkamp, C. H. Lee, A. Szameit, M. Greiter, R. Thomale, Nature
Physics \textbf{16}, 747 (2020)
\end{thebibliography}
\end{document}